# Correlated Insulating States at Fractional Fillings of the WS$_2$/WSe$_2$ Moiré Lattice

Xiong Huang[1,2†], Tianmeng Wang[3†], Shengnan Miao[3†], Chong Wang[4†], Zhipeng Li[3], Zhen Lian[3], Takashi Taniguchi[5], Kenji Watanabe[5], Satoshi Okamoto[6], Di Xiao[4*], Su-Fei Shi[3,7*], Yong-Tao Cui[1,2*]

[1] Department of Materials Science and Engineering, University of California, Riverside, California, 92521, USA

[2] Department of Physics and Astronomy, University of California, Riverside, California, 92521, USA

[3] Department of Chemical and Biological Engineering, Rensselaer Polytechnic Institute, Troy, NY 12180, USA

[4] Department of Physics, Carnegie Mellon University, Pittsburgh, PA 15213, USA

[5] National Institute for Materials Science, 1-1 Namiki, Tsukuba 305-0044, Japan

[6] Materials Science and Technology Division, Oak Ridge National Laboratory, Oak Ridge, Tennessee 37831, USA

[7] Department of Electrical, Computer & Systems Engineering, Rensselaer Polytechnic Institute, Troy, NY 12180, USA

[†] These authors contribute equally to this work.

[*] Correspondence to: dixiao@cmu.edu, shis2@rpi.edu, yongtao.cui@ucr.edu

**Moiré superlattices of van der Waals materials, such as twisted graphene and transitional metal dichalcogenides, have recently emerged as a fascinating platform to study strongly correlated states in two dimensions (2D), thanks to the strong electron interaction in the moiré minibands.[1–19] In most systems, the correlated states appear when the moiré lattice is filled by integer number of electrons per moiré unit cell. Recently, correlated states at fractional fillings of 1/3 and 2/3 holes per moiré unit cell has been reported in the $WS_2$/$WSe_2$ heterobilayer, hinting the long range nature of the electron interaction.[16] In this work, employing a scanning microwave impedance microscopy technique that is sensitive to local electrical properties, we observe a series of correlated insulating states at fractional fillings of the moiré minibands on both electron- and hole-doped sides in angle-aligned $WS_2$/$WSe_2$ hetero-bilayers, with certain states persisting at temperatures up to 120 K. Monte Carlo simulations reveal that these insulating states correspond to ordering of electrons in the moiré lattice with a periodicity much larger than the moiré unit cell, indicating a surprisingly strong and long-range interaction beyond the nearest neighbors. Our findings usher in unprecedented opportunities in the study of strongly correlated states in two dimensions.**

The band theory predicts that a partially filled energy band should produce a metallic state. However, when electron interaction becomes considerable, insulating states can appear at partial fillings of the band. A well-known example is the Mott insulator state at half filling in which strong on-site Coulomb repulsion prevents double occupancy on the same lattice site, resulting in electron localization with exactly one electron per lattice site.[20,21] If we go beyond on-site repulsion, inter-site interactions can lead to even more exotic correlated states at other fractional band fillings, corresponding to commensurate occupation of the lattice sites, such as fractional Chern insulators[22], charge density waves[23], checkerboard and stripe phases[24,25], and Wigner crystals[26–30]. Recently, moiré superlattices based on van der Waals heterostructure of 2D materials emerge as a new playground to engineer correlated states.[1–19] The formation of the periodic moiré structure with a lattice size on the mesoscopic length scale (~10's of nm) results in flat minibands with much reduced kinetic energies, while the Coulomb interaction is strongly enhanced due to the reduced screening in 2D. Correlated states, including Mott insulators and superconductivity, have been observed in various versions of twisted graphene layers at small angles.[1–14] The Mott insulator states appear at fillings of the minibands corresponding to one, two, and three electrons per moiré unit cell. Very recently, it has been shown that the electron interaction is even further strengthened in angle-aligned hetero-bilayers of TMDs[15–19], including $WS_2/WSe_2$[16,17], $WSe_2/WSe_2$[19], and $WSe_2/hBN/WSe_2$[18] systems. The Mott insulator state at the filling of one hole per moiré unit cell[16,17,19] and generalized Wigner crystal states at fillings of 1/3 and 2/3 holes per moiré unit cell[16] have been reported, indicating strong interactions on site and among nearest-neighbors, but these correlated states have only been observed on the hole-doped side. It is intriguing to explore whether these correlated states would form on the electron-doped side, and more importantly, whether the strong interaction can extend beyond nearest neighbors to induce correlated states at more complex commensurate fillings of the underlying moiré lattice.

In this work, we report the observation of a series of correlated insulating states for both electron- and hole-doped regimes in a $WS_2/WSe_2$ moiré heterostructure, including the states at fillings of $n=\pm 1$ (one electron (+1) or hole (-1) per moiré unit cell), and more excitingly, correlated insulating states at a series of fractional fillings including $n=+1/6, \pm 1/4, \pm 1/3, \pm 1/2, \pm 2/3, \pm 3/4, +5/6, \pm 3/2$. Monte Carlo simulations of a Coulomb gas model suggest that they correspond to long-range orderings of electrons in the moiré lattice with spatial patterns of various triangular and stripe phases. In one sample, we find that the transition temperature for $n=\pm1/3$ and $\pm2/3$ can be as high as 120 K, and additional fillings at $n$=-8/9, -5/6, -7/9, +5/9 and +6/7 are also observed, indicating an unexpectedly strong interaction achieved in this sample.

The typical structure of our devices is shown in Fig. 1a. Angle-aligned monolayers of $WS_2$ and $WSe_2$ are encapsulated by hexagonal boron nitride (hBN) flakes, and thin graphite flakes are used as both the electrical contact and bottom gate electrodes (see Methods for details). The lattice constants of $WS_2$ and $WSe_2$ have a ~4% mismatch, which creates a moiré pattern with a periodicity of ~8 nm when the two monolayers are aligned precisely at either 0º or 60º. Multiple minibands should form on both the electron- and hole-doped sides, with the conduction band minimum and valence band maximum locating in different layers according to the type-II band alignment in this hetero-bilayer.[15,17,31,32] Therefore, correlated insulating states can potentially appear on both electron- and hole-doped sides. To probe the insulating states, we employ scanning microwave impedance microscopy (MIM) that is capable of sensing the local resistivity of the sample (Fig. 1b).[33] In MIM, a microwave signal in the frequency range of 1-10 GHz is routed to a sharp metal tip, and the reflected signal is analyzed to extract the imaginary and real parts of the complex tip-sample impedance, which we call MIM-Im and MIM-Re, respectively (see Methods).[34] As the tip voltage oscillates, carriers in the sample move toward and away from the tip to screen the ac electric fields. Such screening capability is characterized by the MIM-Im signal. The MIM-Re signal, on the other hand, characterizes the dissipation generated by the oscillating current induced in the sample. Both channels depend on the sample resistivity as shown in the typical MIM response curves in Fig. 1b. In general, MIM-Im decreases monotonically with increasing resistivity, and there is a finite sensitivity window outside which MIM signals become saturated.

Using MIM, we observe a series of insulating states at fractional filling levels in the WS$_2$/WSe$_2$ moiré lattice on both doping sides. We perform MIM measurement with the tip parked at a fixed spot over the sample while sweeping the gate voltage. A representative data taken at a temperature $T = 3$ K is shown in Fig. 1c. At high doping levels on both electron and hole sides, the sample is highly conductive, resulting in a saturated high MIM-Im signal. As the doping is reduced from either side, the sample becomes more insulating and the MIM-Im signal decreases. In this process, a series of pronounced dips appear in the MIM-Im trace, indicating that the sample goes through several insulating states. The major dips exhibit a structure very similar between electron and hole sides, corresponding to $n = \pm 1/3, \pm 1/2, \pm 2/3, \pm 1, \pm 3/2$, and $+2$. (See Methods for details on the calibration of the filling values, and see Supplementary Information S2 regarding the $n = -2$ state.) There are also several fine features next to the major filling values mentioned above, but their features are not well separated from the major filling values, mostly because the sample resistivity is very high at $T$ = 3 K and the MIM-Im signal becomes saturated at the lower limit. By raising the temperature slightly to reduce the resistivity, the fine features become more pronounced. In Fig. 1d and 1e, we plot traces of MIM-Im vs calibrated filling, $n$, at several temperatures from 3 K to 26 K. From these data, we identify the filling values for several additional states at $n = \pm 1/4$, $\pm 3/4$, $+1/6$, and $+5/6$. Among all the observed states, the $n$=-1/3, -2/3, and -1 states have been reported in recent works, in which the $n$=-1 state was interpretted as a Mott insulator [16,17], and the $n$=-1/3 and -2/3 states as generalized Wigner crystals[16]. According to the type-II band alignment in this system, the states at negative filling values correspond to filling of the valence miniband in the $WSe_2$ layer. Similarly, the ones at positive filling values correspond to filling of the conduction miniband in the $WS_2$ layer. In particular, the $n$=+2 state corresponds to the complete filling of the conduction miniband with two electrons per moiré unit cell.

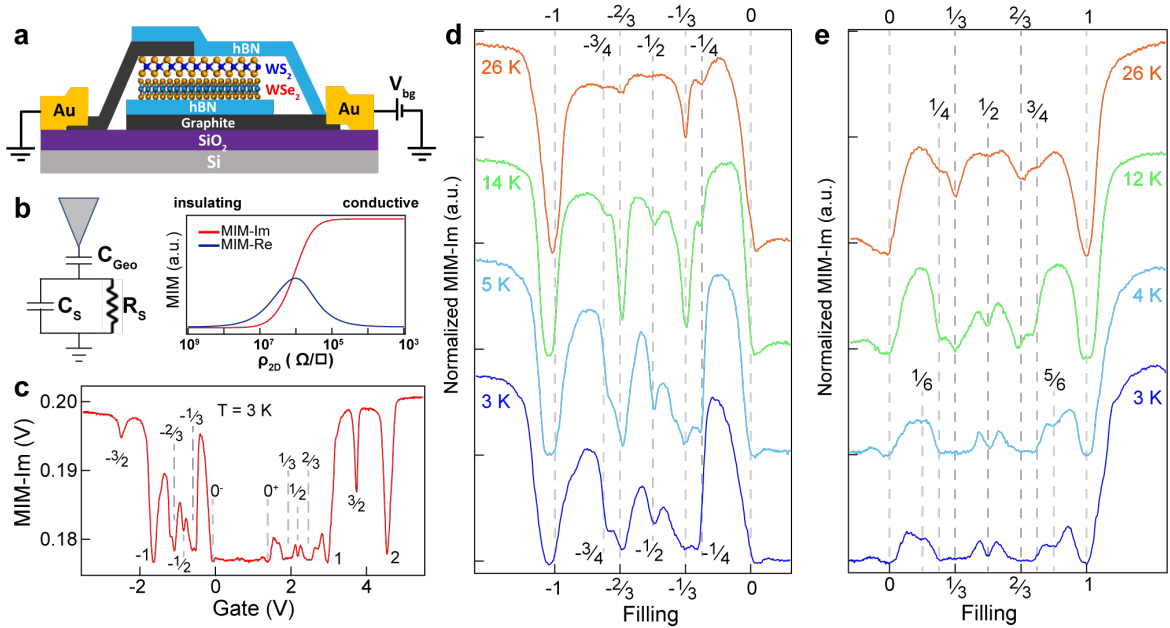

**Figure 1. Spectrum of the correlated insulating states in device D1. a,** Schematics of the device structure. **b,** Effective circuit model for the tip-sample impedance probed by MIM, and a typical response curve showing MIM-Im and MIM-Re as a function of 2D resistivity of the sample. **c,** MIM-Im vs gate curve taken at $T = 3$ K. Dotted lines indicate the nominal positions of fractional fillings at $n=\pm 1/3, \pm 2/3, \pm 1/2$, and the band edges at $0^+$ and $0^-$. **d & e,** MIM-Im vs calibrated filling at selected temperatures for (**d**) hole and (**e**) electron doping sides.

For the $n=\pm 1$ Mott insulator states, the entire moiré lattice is uniformly occupied by exactly one electron/hole per moiré unit cell (Fig. 2a). For the generalized Wigner crystal states at $n=\pm 1/3$, one electron/hole fills in a set of three moiré unit cells so that charges can only occupy every second nearest-neighbor sites, which forms a triangular lattice as illustrated in Fig. 2b. The distance between neighboring occupied sites is $\sqrt{3}a_0$, where $a_0$ is the distance between the nearest-neighbor sites in the moiré lattice. The $n=\pm 2/3$ states are simply complementary to $n=\pm 1/3$ by exchanging occupied and empty sites. All the other fractional fillings should correspond to more complex spatial patterns.

To gain more insights on the ordering patterns in these insulating states at fractional filling $|n| < 1$, we describe our system with a Coulomb gas model on a triangular lattice. For definiteness, we assume $n > 0$. The corresponding Hamiltonian reads

$$H = \frac{1}{2}\sum_{i,j} V_{ij}(n_i - n)(n_j - n)$$

where $V_{ij}$ is the inter-site repulsion and $n_i = 0$ or 1 is the occupation number of site *i*. We have subtracted the average charge density, which is the filling fraction *n*, to ensure charge neutrality. The Hamiltonian above is manifestly particle-hole symmetric, i.e., the state at filling faction 1-*n* can be obtained from that at *n* by swapping occupied and empty sites. Here we have neglected the kinetic energy completely. Since the period of the moiré lattice is relatively large (~ 8 nm), long-range hopping should be exponentially suppressed. To capture the essence of the interaction effect, we consider the unscreened Coulomb interaction, $V_{ij} = 1/|\mathbf{R}_i - \mathbf{R}_j|$. Note that in a 2D device with a nearby metallic gate, the Coulomb interaction is in general screened and can be approximated by $V(r) = 1/r - 1/\sqrt{r^2 + 4D^2}$, where *D* is the distance from the gate to the device. The unscreened Coulomb potential is therefore appropriate for large *D*.

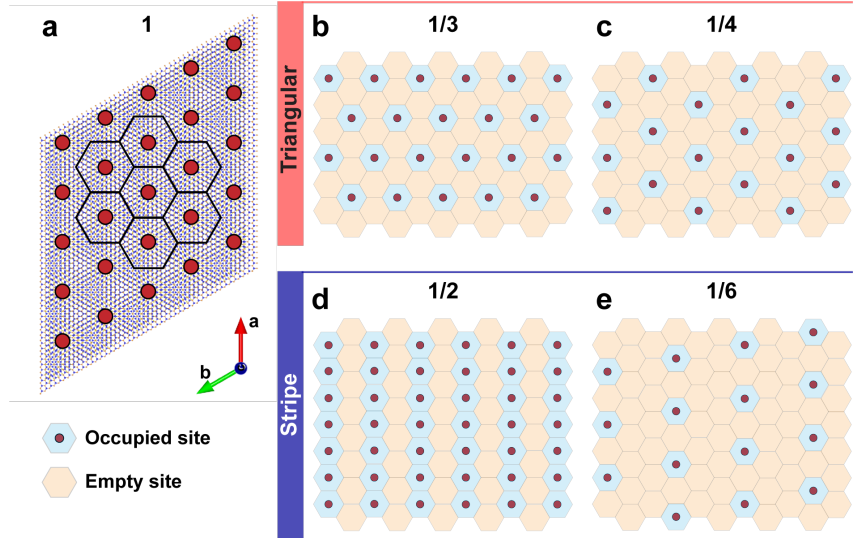

**Figure 2. Ordering patterns at fractional fillings of the moiré lattice. a,** Schematics of the triangular lattice for the $WS_2/WSe_2$ moiré superlattice, filled by one carrier per moiré unit cell. **b & c,** The patterns for the triangular lattice formed at fractional fillings of *n*=1/3 and 1/4. **d & e,** The patterns for the stripe phase formed at fractional fillings of *n*=1/2 and 1/6.

We perform Monte Carlo (MC) simulations based on this simple Coulomb gas model to search through the filling fractions ranging from 0 to 1/2 (since filling fractions $n$ and 1-$n$ are particle-hole symmetric) and identify the charge ordering patterns for the observed filling fractions (see Methods and Supplementary Information S5 for more details on the simulation procedure), which can be grouped into two categories, triangular and stripe phases, as shown in Fig. 2. (More filling patterns for another device will be presented in Fig. 4.) The triangular phases include $n$=1/3 and 1/4 (Fig. 2b-2c). They match the observed filling values at $n$=±1/3, ±2/3, ±1/4 and ±3/4. These states can be considered as generalized Wigner crystals[27], as a result of the competition between the long-range Coulomb interaction and the confining potential in the moiré lattice. They preserve the $C_3$ rotational symmetry of the original moiré lattice but break its translational symmetry. In general, the triangular lattice can form for any filling value $n$=1/$p$, where $\sqrt{p}a_0$ matches the distance between two nearby neighbor sites. For example, the 1/3 state corresponds to the second nearest-neighbor (NN) separation, and the 1/4 state corresponds to the third NN.

On the other hand, the stripe phases further break the $C_3$ rotational symmetry and should occur for commensurate fractions at higher densities. We find that the stripe phase is a robust ground state configuration for $n$=1/2 and 1/6, which corresponds to the $n$=±1/2, +1/6, and +5/6 observed in our experiment. The $n$=±3/2 states can be understood as adding carriers on top of a uniformly occupied $n$=±1 states, and therefore, they should be in the same stripe phase pattern as $n$=1/2, if we ignore the specific occupation site within the moiré unit cell. Each stripe consists of a few rows, and a fraction of the sites are occupied according to the filling value. Neighboring stripes have a translational shift along the stripe direction. The combination of intra-stripe occupation and inter-stripe shift minimizes the overall interaction energy of the system. Transitions into the stripe phase at $n$=1/2 and 1/6 have well-defined ordering temperatures, indicating their robustness against perturbations.

Next we study the temperature dependence of the insulating states to examine the melting of the charge ordering. Figure 3a plots the MIM-Im vs gate voltage traces from 3 K up to 160 K. The states at $n$=±1 can be resolved at temperatures up to 120 K – 160 K, which corresponds to energy gaps of ~10 meV, consistent with recent studies[17]. The fractional fillings at $n$=±1/3, ±2/3, ±1/2, ±1/4, ±3/4, +3/2, and +2 disappear at around 30-40 K. The $n$=-3/2 state disappears at ~10 K, while the $n$=+1/6 and +5/6 states disappear at 4-6 K. Figure 3b plots the temperature ranges in which each filling can be resolved by MIM. The disappearance of the insulating states suggests that the carriers are no longer localized but can easily hop to nearby empty sites. We can then associate this characteristic temperature with the melting of the charge orders seen in the MC simulations at elevated temperatures. As a comparison, Fig. 3c plots the ordering temperatures obtained through our MC simulation, which match well the experimental data.

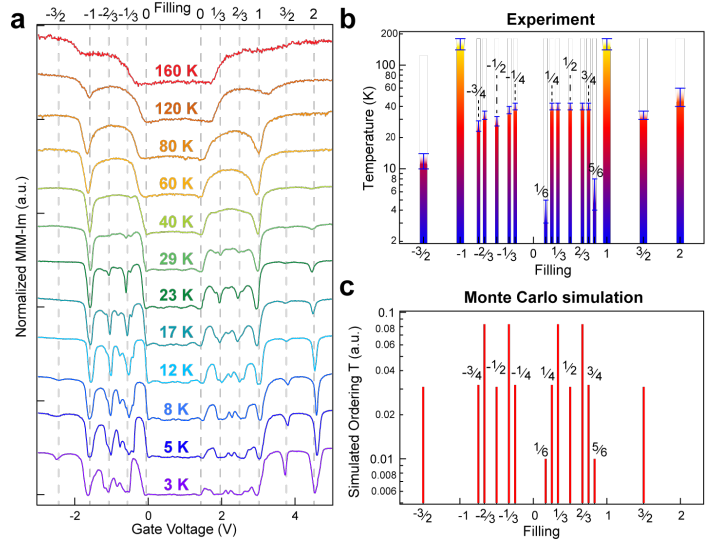

**Figure 3. Temperature dependence of the correlated insulating states in device D1. a,** MIM-Im vs gate curves taken at temperatures from 3 K to 140 K. **b,** The temperature range in which the correlated insulating states can be resolved by MIM. **c,** The ordering temperatures obtained from Monte Carlo simulations, for the filling fractions corresponding to the experimentally observed values.

In device D2, we observe behaviors that indicate an even longer-range and stronger interaction than that in D1. Figure 4a plots the MIM-Im spectrum as a function of calibrated filling measured at $T = 2.8$ K. Compared to D1, we observe several new filling fractions including $n=-8/9, -5/6, -7/9, +5/9$, and $+6/7$. Among these fillings, the $n=-5/6$ state is the counterpart to the $n=+5/6$ state, which has a stripe pattern. The $n= +6/7$ and $-8/9$ states are found to have triangular patterns by our MC simulation, plotted as $1/7$ and $1/9$, respectively, in Fig. 4b. The $1/7$ filling corresponds to the fourth NN separation and the $1/9$ corresponds to the fifth NN, much longer range than the triangular lattice states observed in D1. The $n= +5/9$ and $-7/9$ states are two exceptions that our simulations do not obtain well-defined transitions to the ground states, due to many energetically close configurations (see Supplementary Information S5). The $n=\pm1/3, \pm2/3$, and $\pm1$ states can be resolved at temperatures up to $100 – 120$ K (see Fig. 4c), while all other fractional fillings disappear at around 20-30 K. The transition temperature for $n=\pm1/3$ and $\pm2/3$ is much enhanced compared to our other samples as well as earlier studies[16], signaling a stronger interaction strength in D2 which corroborates the observation of rich filling fractions. While the mechanism for the enhancement in D2 is not yet clear, we suspect that local strain or inhomogeneities could play a role (see Supplementary Information S3 for more data taken at nearby spots). Current generation of moiré devices are known to have large spatial inhomogeneities.[35] The spatial resolution of our MIM technique, on the order of 100 nm, allows us to detect local features that exhibit strong interactions.

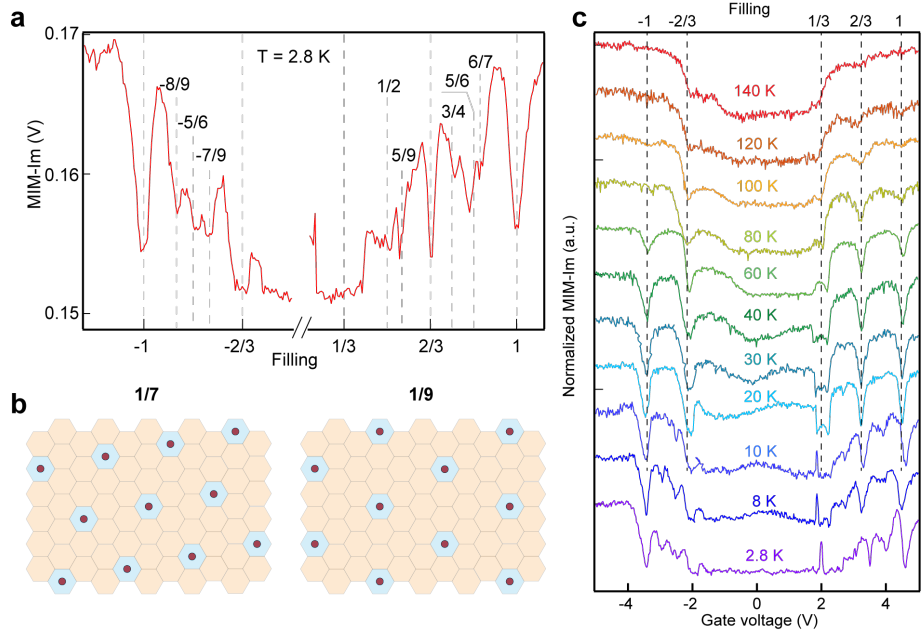

**Figure 4. Correlated insulating states in device D2. a,** MIM-Im vs calibrated filling at $T = 2.8$ K. Dotted lines indicate the nominal positions of fractional fillings as labelled. **b,** The spatial filling patterns for $n=1/7$ and $1/9$, obtained through MC simulation. **c,** Temperature dependence of the MIM-Im spectrum from 2.8 K to 140 K. Dotted lines indicated the positions for the $n=+1/3, \pm2/3$, and $\pm1$ determined from the high temperature data.

In summary, our observation of correlated states at multiple filling fractions, especially with the prospect of high transition temperatures for certain states, highlights the potential to realize strong long-range electron interactions in the TMD moiré heterostructure. Together with its gate tunability and flexibility to integrate with other 2D materials, the TMD moiré heterostructure offers a promising playground to further explore exotic correlated states such as fractional quantum Hall state[36,37], superconductivity[38], and quantum spin liquid[39] in 2D systems.

*Note added:* During the revision of our manuscript, we learned about similar results on the observation of correlated states at fractional fillings in the $WS_2$/$WSe_2$ moiré lattice.[40,41]

**Methods:**

Heterostructure device fabrication: We use a layer-by-layer stacking method to fabricate the $WS_2$/$WSe_2$ heterostructures. The bulk crystals of $WS_2$ and $WSe_2$ are acquired from HQ Graphene. We first exfoliate monolayer $WS_2$, monolayer $WSe_2$, few-layer graphene and thin hBN flakes onto PDMS layers on glass slides, respectively. For aligned heterostructures, we choose exfoliated monolayers with sharp edges, whose crystal axes are further confirmed by second harmonic generation (SHG) measurements. We then mount the silicon substrate with a 285 nm thermal oxide layer on a rotational stage and clamp the glass slide with thin flakes to another three-dimensional (3D) stage. We adjust the 3D stage to control the distance between substrates and thin flakes, and stack the thin graphite bottom gate, hBN flake, WSe$_2$ monolayer, WS$_2$ monolayer, the thin graphite contact, and the second hBN flake in sequence onto the pre-patterned Au electrodes on Si substrates. We fine adjust the angle of the rotational stage (accuracy of 0.02°) under a microscope objective to stack the monolayer WS$_2$ on the monolayer WSe$_2$, ensuring a near-zero twist angle between the two flakes. The final constructed device is annealed at 140 °C for 6 hours in a vacuum chamber. The pre-patterned Au contact electrodes are fabricated through standard electron-beam lithography and ebeam evaporation processes.

Microwave impedance microscopy measurements: The MIM measurement is performed on a homebuilt cryogenic scanning probe microscope platform. A small microwave excitation of ~0.1 μW at a fixed frequency around 10 GHz is delivered to a chemically etched tungsten tip mounted on a quartz tuning fork. The reflected signal is analyzed to extract the demodulated output channels, MIM-Im and MIM-Re, which are proportional to the imaginary and real parts of the admittance between the tip and the sample, respectively. To enhance the MIM signal quality, the tip on the tuning fork is excited to oscillate at a frequency of ~32 kHz with an amplitude of ~8 nm. The resulting oscillation amplitudes of MIM-Im and MIM-Re are then extracted using a lock-in amplifier to yield d(MIM-Im)/d$z$ and d(MIM-Re)/d$z$, respectively. The d(MIM)/d$z$ signals are free of fluctuating backgrounds, thus enabling more quantitative analysis, while their behavior is very similar to that of the standard MIM signals. In this paper we simply refer to d(MIM)/d$z$ as the MIM signal.

Determination of the alignment angle of the heterobilayer: The twist angle between WS$_2$ and WSe$_2$ flakes is measured by SHG (see Supplementary Information S1). It is 1.1° ± 0.3° off 60° alignment in device D1 and 2.2°±0.3° off 0° alignment in device D2.

The angle can be also estimated from the charge density corresponding to the $n$=1 state, $n_0$, calculated using a parallel capacitor model: $n_0 = \epsilon_0 \epsilon_r \Delta V_g / ed$ where the $\epsilon_0$ is the vacuum permittivity, $\epsilon_r = 3 \sim 4$ is the dielectric constant of hBN, $d$ is the thickness of the bottom hBN, and $e$ the electron charge. The gate voltage $\Delta V_g$ is determined from the filling fraction calibration. For device D1, $d$ = 17.4 nm and $\Delta V_g$ = 1.50 V, we obtain $n_0$ = 1.4~1.9 × 10$^{12}$ cm$^{-2}$, corresponding to a twist angle < 0.7° off 60° alignment. For device D2, $d$ = 23.3 nm and $\Delta V_g$ = 3.84 V, we obtain $n_0$ = 2.7~3.6 × 10$^{12}$ cm$^{-2}$, corresponding to a twist angle of 1.6°~2.4° off 0° alignment. Both are consistent with the SHG results.

Determination of the filling fractions: For both the electron and hole sides, the positions of the |$n$|=1/3, 2/3, and 1 states are identified first. By assuming a linear relation between gate voltage and carrier density, the calibration from gate voltage to filling values can be completely determined by fitting these three data points with a linear line on the respective doping side. Then for each observed feature, its filling value in the decimal format is calculated based on the calibration and then converted to the closest simple fractional format. (In D2, only |$n$|=2/3 and 1 are used for calibration. See Supplementary Information S6 for more information.)

Monte Carlo simulations: The Coulomb lattice gas model is simulated with the Markov chain Monte Carlo method. Parallel tempering[40] is used to avoid critical slowing down around the transition temperature and Ewald summation[41] is used to capture the long-range interaction of Coulomb potential. Most of the simulations are performed in a 12 × 12 lattice, except for filling fraction 1/7, which is performed on a 14 × 14 lattice. 48 × 48 lattice is also used to examine the formation of domains (see Supplementary Information S5). 72 replicas have been used and 10 million sweeps are performed for each replica. The temperature list for the replicas is a geometric series starting from $k_B T_1 = 0.01$ with $T_{n+1}/T_n = 1.032$, where $k_B T$ is normalized with respect to the Coulomb interaction between nearest-neighbor sites. In each sweep, simple Metropolis update is used and exchanges between neighboring replicas are attempted every 10 sweeps. The second half of the sweeps are used to calculate the specific heat. The transition temperatures are estimated from the $C_v - T$ curve together with the Fourier spectrum of the charge pattern.

**References**

1. Bistritzer, R. & MacDonald, A. H. Moire bands in twisted double-layer graphene. *Proc. Natl. Acad. Sci.* **108**, 12233–12237 (2011).
2. Cao, Y. *et al.* Correlated insulator behaviour at half-filling in magic-angle graphene superlattices. *Nature* **556**, 80–84 (2018).
3. Cao, Y. *et al.* Unconventional superconductivity in magic-angle graphene superlattices. *Nature* **556**, 43–50 (2018).
4. Shen, C. *et al.* Correlated states in twisted double bilayer graphene. *Nat. Phys.* **16**, 520–525 (2020).
5. Cao, Y. *et al.* Tunable correlated states and spin-polarized phases in twisted bilayer–bilayer graphene. *Nature* **583**, 215–220 (2020).
6. Liu, X. *et al.* Tunable spin-polarized correlated states in twisted double bilayer graphene. *Nature* **583**, 221–225 (2020).
7. Chen, G. *et al.* Evidence of a gate-tunable Mott insulator in a trilayer graphene moiré superlattice. *Nat. Phys.* **15**, 237–241 (2019).
8. Chen, G. *et al.* Signatures of Gate-Tunable Superconductivity in Trilayer Graphene/Boron Nitride Moire Superlattice. *Nature* **572**, 215–219 (2019).
9. Chen, S. *et al.* Electrically tunable correlated and topological states in twisted monolayer–bilayer graphene. *Nat. Phys.* (2020). https://doi.org/10.1038/s41567-020-01062-6
10. Polshyn, H. *et al.* Nonvolatile switching of magnetic order by electric fields in an orbital Chern insulator. (2020) arXiv:2004.11353.
11. Shi, Y. *et al.* Tunable van Hove Singularities and Correlated States in Twisted Trilayer Graphene. (2020) arXiv:2004.12414.
12. Sharpe, A. L. *et al.* Emergent ferromagnetism near three-quarters filling in twisted bilayer graphene. *Science* **365**, 605–608 (2019).
13. Serlin, M. *et al.* Intrinsic quantized anomalous Hall effect in a moiré heterostructure. *Science* **367**, 900–903 (2020).
14. Chen, G. *et al.* Tunable correlated Chern insulator and ferromagnetism in a moiré superlattice. *Nature* **579**, 56–61 (2020).
15. Wu, F., Lovorn, T., Tutuc, E. & MacDonald, A. H. H. Hubbard Model Physics in Transition Metal Dichalcogenide Moiré Bands. *Phys. Rev. Lett.* **121**, 26402 (2018).
16. Regan, E. C. *et al.* Mott and generalized Wigner crystal states in WSe2/WS2 moiré

superlattices. *Nature* **579**, 359–363 (2020).
17. Tang, Y. *et al.* Simulation of Hubbard model physics in WSe2/WS2 moiré superlattices. *Nature* **579**, 353–358 (2020).
18. Shimazaki, Y. *et al.* Strongly correlated electrons and hybrid excitons in a moiré heterostructure. *Nature* **580**, 472–477 (2020).
19. Wang, L. *et al.* Correlated electronic phases in twisted bilayer transition metal dichalcogenides. *Nat. Mater.* **19**, 861–866 (2020).
20. Mott, N. F. The Basis of the Electron Theory of Metals, with Special Reference to the Transition Metals. *Proc. Phys. Soc. Sect. A* **62**, 416–422 (1949).
21. Imada, M., Fujimori, A. & Tokura, Y. Metal-insulator transitions. *Rev. Mod. Phys.* **70**, 1039–1263 (1998).
22. Sheng, D. N., Gu, Z.-C., Sun, K. & Sheng, L. Fractional quantum Hall effect in the absence of Landau levels. *Nat. Commun.* **2**, 389 (2011).
23. Grüner, G. The dynamics of charge-density waves. *Rev. Mod. Phys.* **60**, 1129–1181 (1988).
24. Emery, V. J., Kivelson, S. A. & Tranquada, J. M. Stripe phases in high-temperature superconductors. *Proc. Natl. Acad. Sci.* **96**, 8814–8817 (1999).
25. Hoffman, J. E. A Four Unit Cell Periodic Pattern of Quasi-Particle States Surrounding Vortex Cores in Bi2Sr2CaCu2O8+delta. *Science* **295**, 466–469 (2002).
26. Wigner, E. On the interaction of electrons in metals. *Phys. Rev.* **46**, 1002–1011 (1934).
27. Hubbard, J. Generalized wigner lattices in one dimension and some applications to tetracyanoquinodimethane (TCNQ) salts. *Phys. Rev. B* **17**, 494–505 (1978).
28. Grimes, C. C. & Adams, G. Evidence for a Liquid-to-Crystal Phase Transition in a Classical, Two-Dimensional Sheet of Electrons. *Phys. Rev. Lett.* **42**, 795–798 (1979).
29. Andrei, E. Y. *et al.* Observation of a Magnetically Induced Wigner Solid. *Phys. Rev. Lett.* **60**, 2765–2768 (1988).
30. Shapir, I. *et al.* Imaging the electronic Wigner crystal in one dimension. *Science* **364**, 870–875 (2019).
31. Jin, C. *et al.* Observation of moiré excitons in WSe2/WS2 heterostructure superlattices. *Nature* **567**, 76–80 (2019).
32. Zhang, Z. *et al.* Flat bands in twisted bilayer transition metal dichalcogenides. *Nat. Phys.* **16**, 1093–1096 (2020).
33. Lai, K., Kundhikanjana, W., Kelly, M. & Shen, Z. X. Modeling and characterization of a cantilever-based near-field scanning microwave impedance microscope. *Rev. Sci. Instrum.* **79**, 063703 (2008).
34. Cui, Y.-T., Ma, E. Y. & Shen, Z.-X. Quartz tuning fork based microwave impedance microscopy. *Rev. Sci. Instrum.* **87**, 063711 (2016).
35. Chu, Z. *et al.* Nanoscale Conductivity Imaging of Correlated Electronic States in WSe2/WS2 Moiré Superlattices. *Phys. Rev. Lett.* **125**, 186803 (2020).
36. DaSilva, A. M., Jung, J. & MacDonald, A. H. Fractional Hofstadter States in Graphene on Hexagonal Boron Nitride. *Phys. Rev. Lett.* **117**, 036802 (2016).
37. Wu, F. & MacDonald, A. H. Moiré assisted fractional quantum Hall state spectroscopy. *Phys. Rev. B* **94**, 241108 (2016).
38. Slagle, K. & Fu, L. Charge Transfer Excitations, Pair Density Waves, and Superconductivity in Moiré Materials. (2020) arXiv:2003.13690.
39. Grover, T., Trivedi, N., Senthil, T. & Lee, P. A. Weak Mott insulators on the triangular

lattice: Possibility of a gapless nematic quantum spin liquid. *Phys. Rev. B* **81**, 245121 (2010).
40. Xu, Y. *et al.* Abundance of correlated insulating states at fractional fillings of WSe2/WS2 moiré superlattices. (2020) arXiv:2007.11128.
41. Jin, C. *et al.* Stripe phases in WSe2/WS2 moiré superlattices. (2020) arXiv:2007.12068.
42. Swendsen, R. H. & Wang, J.-S. Replica Monte Carlo Simulation of Spin-Glasses. *Phys. Rev. Lett.* **57**, 2607–2609 (1986).
43. Toukmaji, A. Y. & Board, J. A. Ewald summation techniques in perspective: a survey. *Comput. Phys. Commun.* **95**, 73–92 (1996).

**Acknowledgments**
We thank Dr. Dongxue Chen, Li Yan, Lei Ma, and Kang Li for help on the device fabrication. We are grateful to Prof. Robert Swendsen and Prof. Michael Widom for their help with the Monte Carlo simulation. C.W. and D.X. thank Prof. Wenhui Duan for providing part of the computational resources. X.H. and Y.-T.C. acknowledge support from the Hellman Fellowship award and the seed fund from SHINES, an EFRC funded by DOE BES under award number SC0012670. S. M., Z. Li and S.-F. S. acknowledge support by AFOSR through Grant FA9550-18-1-0312. T.W. and S.-F.S. acknowledge support from ACS PRF through Grant 59957-DNI10. Z. Lian and S.-F.S. acknowledge support from NYSTAR through Focus Center-NY–RPI Contract C150117. The device fabrication was supported by the Micro and Nanofabrication Clean Room (MNCR) at Rensselaer Polytechnic Institute (RPI). S.-F. S. also acknowledges the support from NSF through Career Grant DMR-1945420. The research by S.O. was supported by the U. S. Department of Energy, Office of Science, Basic Energy Sciences, Materials Sciences and Engineering Division. D.X. is supported by the Department of Energy, Basic Energy Sciences, Grant No. DE-SC0012509. K.W. and T.T. acknowledge support from the Elemental Strategy Initiative conducted by the MEXT, Japan and the CREST (JPMJCR15F3). We acknowledge computing time provided by BRIDGES at the Pittsburgh supercomputer center (Award No. TG-DMR190080) under the Extreme Science and Engineering Discovery Environment (XSEDE) supported by NSF (ACI-1548562).

**Author contributions**
X.H., T.W., S.M., and C.W. contribute equally to this work. S.-F.S. and Y.-T.C. initiated the research. T.W., S.M., Z. Li and Z. Lian fabricated the heterostructure devices. X.H. performed the MIM measurements. Y.-T.C., S.-F.S., D.X., S. M., T. W., C.W., and X.H. analyzed the data. C.W., S.O. and D.X. performed numerical simulations. Y.-T.C., S.-F. S., and D.X. wrote the manuscript with inputs from all authors.

Supplementary Information for

**Correlated Insulating States at Fractional Fillings of the WS$_2$/WSe$_2$ Moiré Lattice**

Xiong Huang[1,2†], Tianmeng Wang[3†], Shengnan Miao[3†], Chong Wang[4†], Zhipeng Li[3], Zhen Lian[3], Takashi Taniguchi[5], Kenji Watanabe[5], Satoshi Okamoto[6], Di Xiao[4*], Su-Fei Shi[3,7*], Yong-Tao Cui[1,2*]

† These authors contribute equally to this work.

* Correspondence to: dixiao@cmu.edu, shis2@rpi.edu, yongtao.cui@ucr.edu

**S1. Device information**

**S2. Additional MIM data in device D1**

**S3. Additional MIM data in device D2**

**S4. MIM data in device D3**

**S5. Monte Carlo simulation**

**S6. Assignment of the filling fractions in device D2**

## S1. Device information

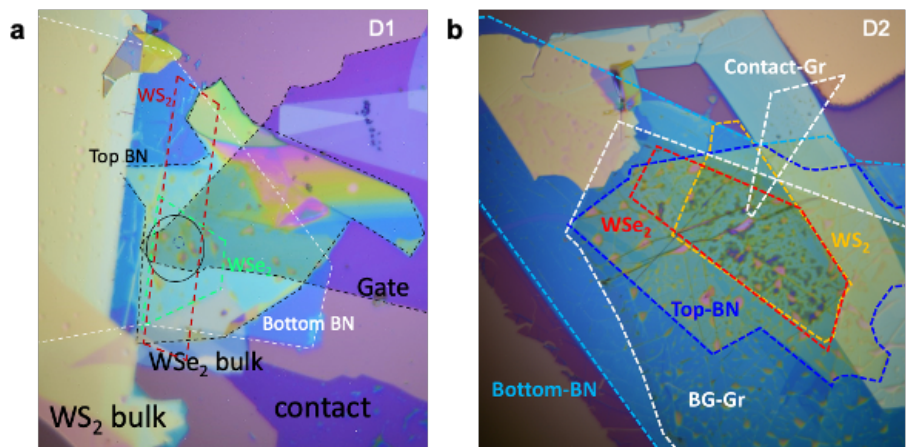

Figure S1. Optical microscope images of devices (a) D1 and (b) D2 with different flakes outlined.

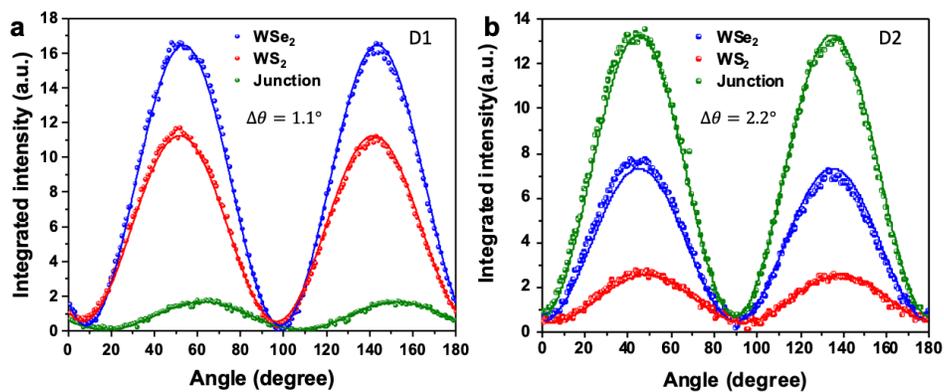

Figure S2. Angular dependence of the SHG signal in (a) device D1 and (b) device D2.

## S2. Additional MIM data in device D1

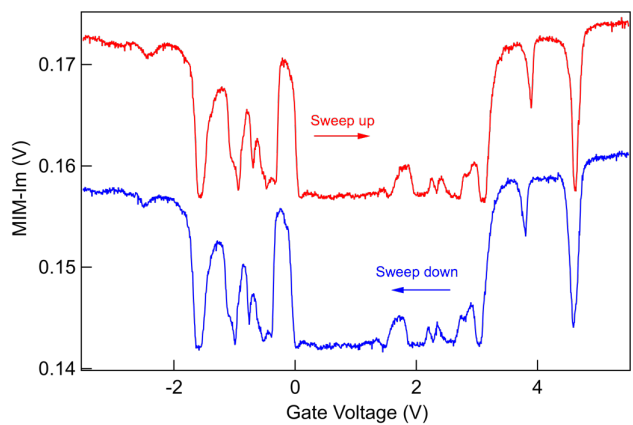

Figure S3. MIM-Im vs gate voltage traces for both sweeping directions in device D1 at $T = 4$ K.

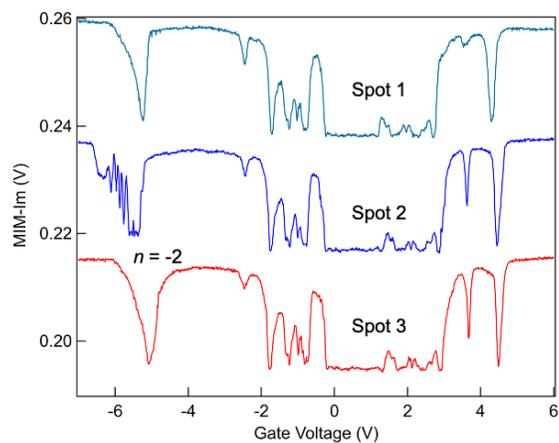

Figure S4. MIM-Im vs gate voltage traces with extended gate range from -7 V to 6 V taken at $T = 3$ K. The feature near -5 V likely correponds to $n = -2$ state. However, due to poor electrical contact at high hole doping, this feature is not stable and its gate position is not repeatable at different spots.

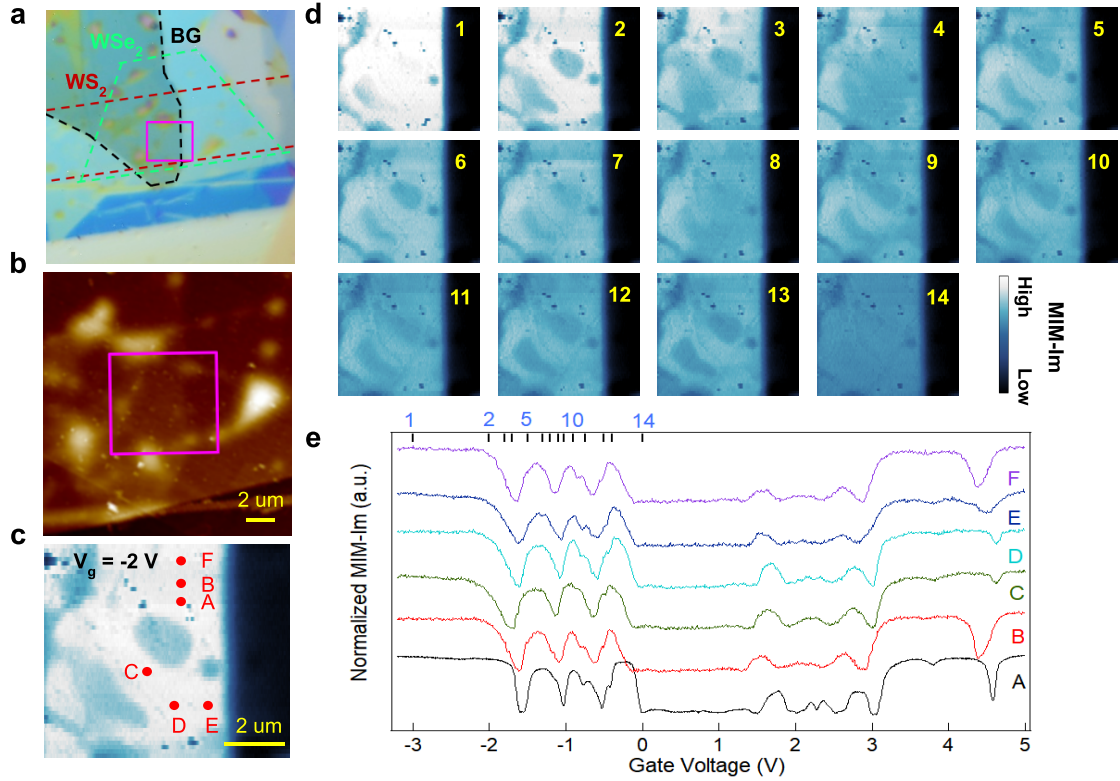

Figure S5. Spatial uniformity of sample conductivity. (a) Optical image and (b) Atomic force microscopy (AFM) image of device D1. (c) MIM-Im image at $V_g = -2$ V scanned over the region marked by the rectangles in (a) and (b). (d) MIM-Im images taken at different gate voltages marked along the top axis in (e). (e) MIM-Im vs gate traces taken at spots A-F as indicated in (c). All MIM data are taken at $T = 10$ K.

## S3. Additional MIM data in device D2

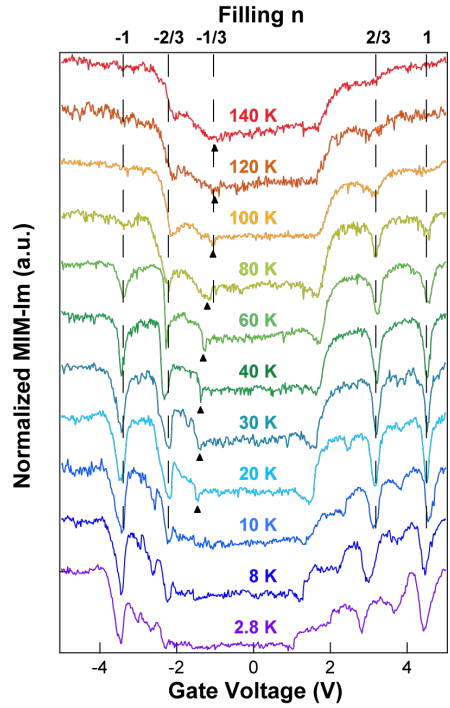

Figure S6. Temperature dependence of MIM-Im vs gate, for gate sweeping down from 5 V to -5 V, in device D2. The black markers indicate the positions of the $n$=-1/3 state, which can be resolved at temperatures of 20 K and above.

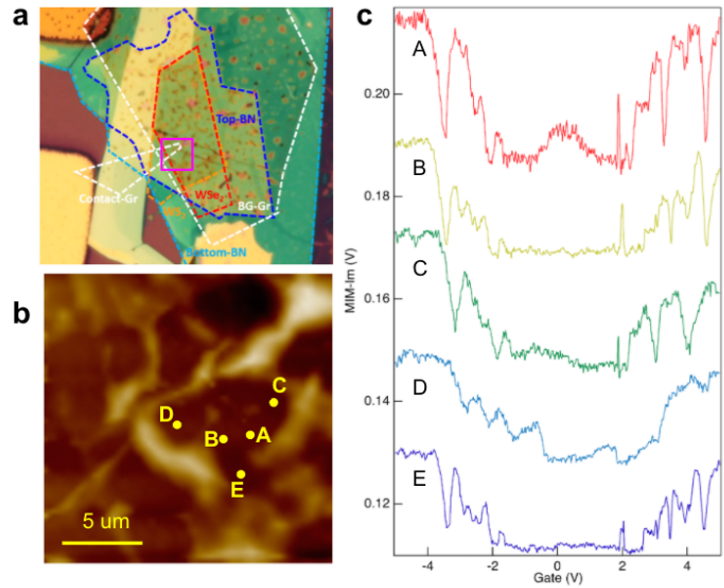

Figure S7. Spatial variation of the MIM-Im spectrum in device D2. (a) Optical microscope of D2. (b) AFM scan around the area indicated by the solid square in (a). (c) MIM spectra taken at spots A-E as indicated in (b). All data taken at $T$ = 2.8 K.

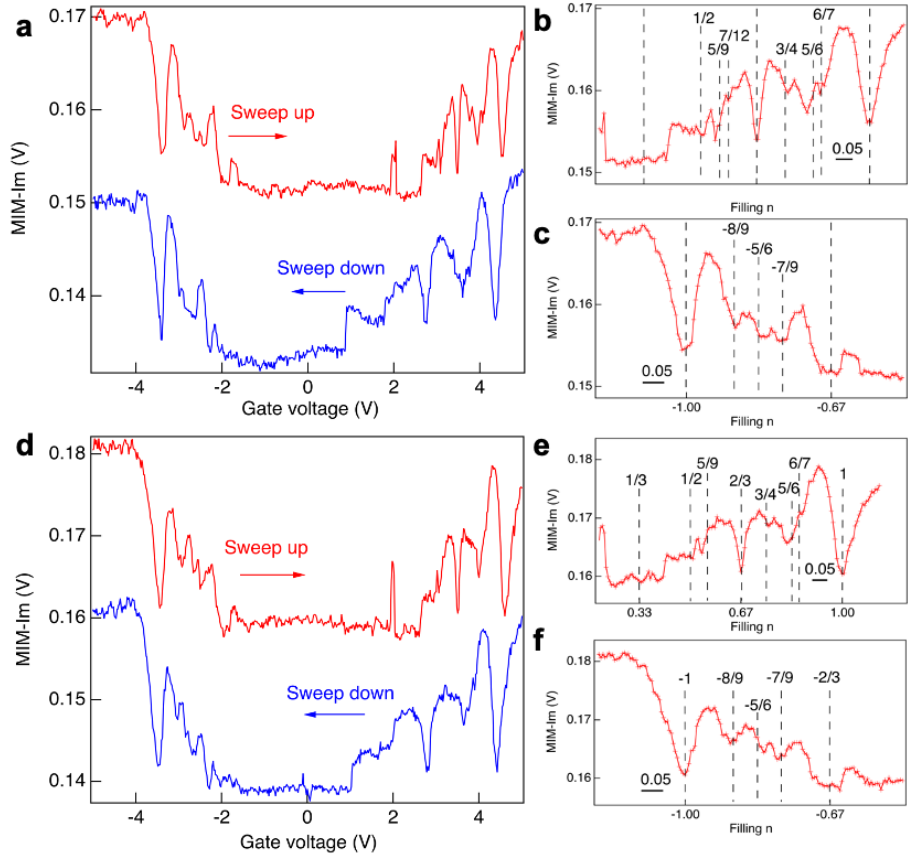

Figure S8 (a) Sweep up and down traces at spot E in Fig. S7b. (b & c) Zoom-in portions of the sweep up curve in (a) with calibrated filling values for (b) electron and (c) hole doping sides. (d) Sweep up and down traces at spot B in Fig. S7b. (e & f) Zoom-in portions of the sweep up curve in (d) with calibrated filling values for (e) electron and (f) hole doping sides. All data are taken at $T = 2.8$ K.

## S4. MIM data in device D3

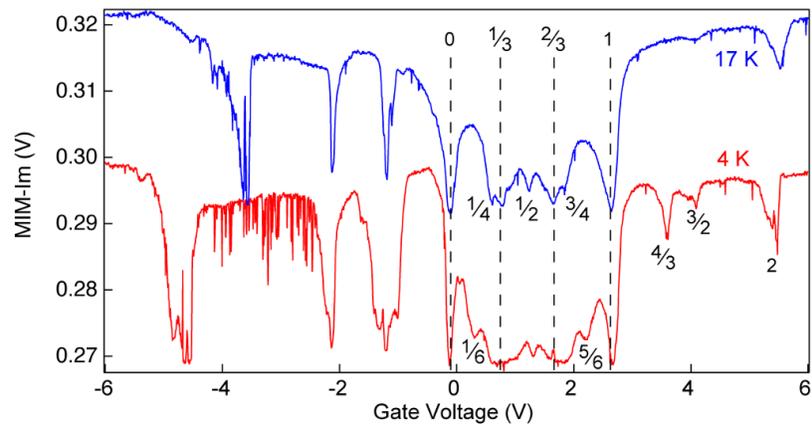

Figure S9. The MIM-Im vs gate voltage traces in device D3 with a twist angle of ~0º, at $T = 4$ K and 17 K. The filling fractions on the electron doping side is similar to those observed in D1. The features on the hole doping side are not very stable due to poor electrical contact.

## S5. Monte Carlo simulation

The lattice Coulomb gas model is simulated with parallel tempering Monte Carlo method as described in the Methods section of the main text. For different fillings, we plot in Fig. S10 the specific heat as a function of the temperature. The transition temperatures, listed in Tab. S1, are estimated from the $C_v - T$ curve together with the Fourier spectrum of the charge pattern.

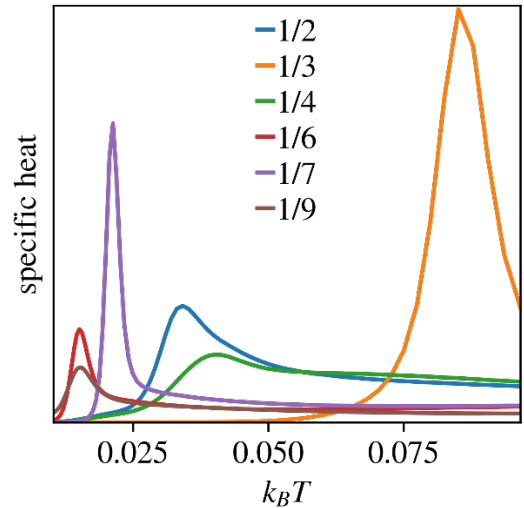

Figure S10. Specific heat for different fillings with respect to the temperature scale, $k_B T$, which is normalized by the nearest neighbor Coulomb interaction.

| Filling | $k_B T_c$ |
|---------|-----------|
| 1/2     | 0.034     |
| 1/3     | 0.085     |
| 1/4     | 0.040     |
| 1/6     | 0.015     |
| 1/7     | 0.021     |
| 1/9     | 0.015     |

Table S1. Transition temperatures for different fillings.

To visualize the transition around the critical temperature. We select 8 configurations from the simulations of 1/7 filling, on which we perform Fourier transformations. The absolute values of the 8 Fourier amplitudes of are added and plotted in the reciprocal space in Fig. S11. It is clear the Fourier amplitudes changes abruptly around $T_c$, which justifies our approach of identifying the peak in the specific heat as the transition temperature.

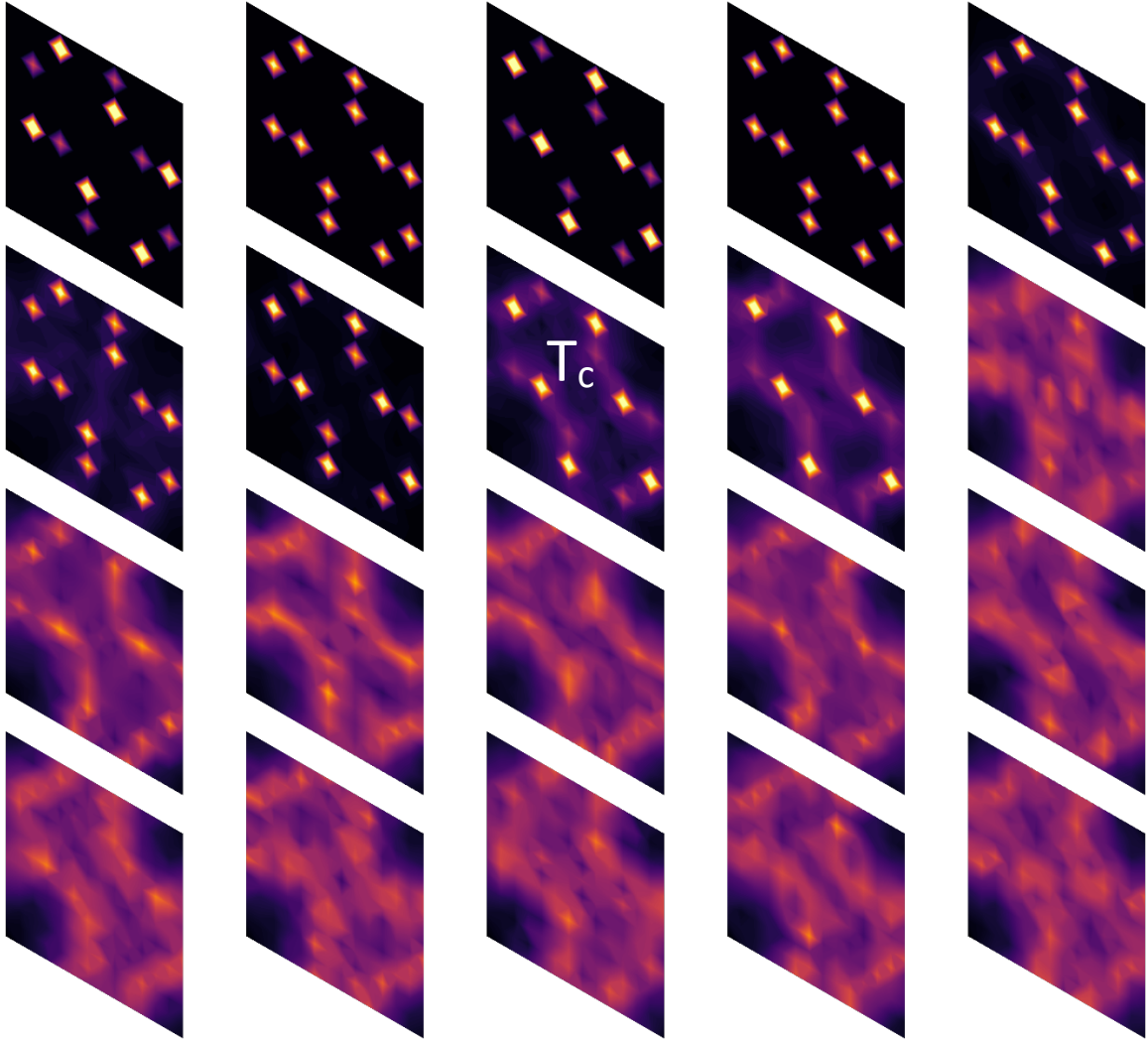

Figure S11. The Fourier transformed configurations in the reciprocal space as $k_B T$ varies for filling 1/7. The lowest $k_B T$ is 0.017; the highest $k_B T$ is 0.031; $k_B T_c = 0.021$ is labeled in the picture.

We have also performed Monte Carlo simulations on larger lattices. In Figs. S12 and S13, we plot ordering patterns for simulations on a $48 \times 48$ lattice. On this lattice, it is difficult to obtain the exact ground state. Instead, the pattern is dominated by large domains, separated by defects or domain walls.

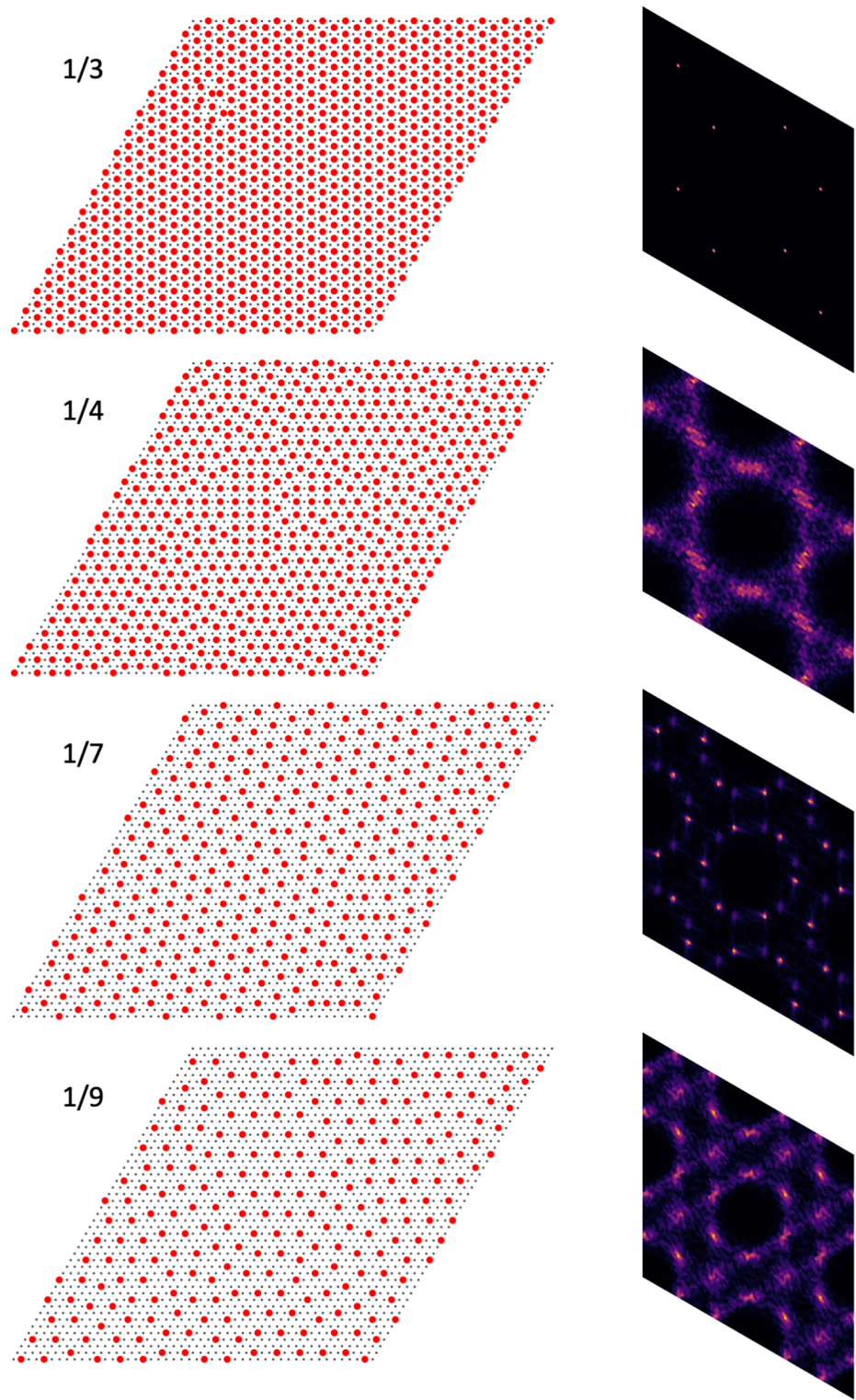

Figure S12. Ordering patterns simulated on a 48 by 48 lattice for *n*=1/3, 1/4, 1/7, and 1/9. The column on the right plots the FFT image of the ordering pattern.

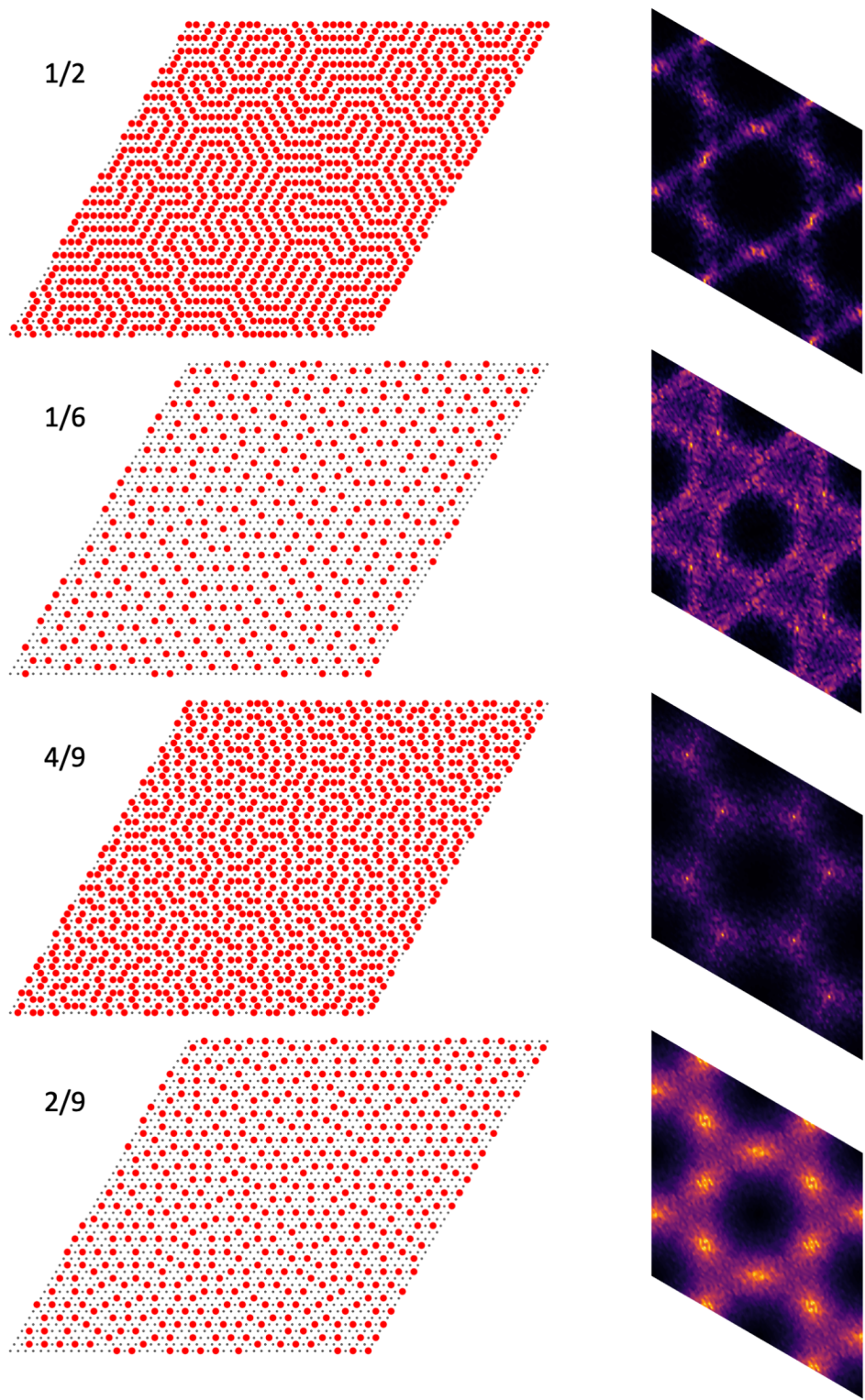

Figure S13. Ordering patterns simulated on a 48 by 48 lattice for $n$=1/2, 1/6, 4/9, and 2/9. The column on the right plots the FFT image of the ordering pattern.

**S6. Assignment of the filling fractions in device D2.**

The assignment of the fractional filling values is described in the Methods section of the main text. The main source of uncertainty comes from the linearity between the gate voltage and density. In device D1, the hysteresis is small and the filling determination is less an issue. Device D2 has a considerable hyesteresis. Since it shows rich features which indicate a stronger interaction strength, we would like to discuss the uncertainties on the filling assignment and the implications on the real space ordering patterns.

We assume that at high gate voltages ($|V_g|$>2 V) in device D2, the charge density has a linear relation with the gate voltage, which is the basis to calculate all the filling values. This assumption can be justified in the following way. On the electron side, 1/3, 2/3 and 1 are all clearly resolved in the 2.8 K data (see Fig. 4a in the main text). We use the 2/3 and 1 positions to determine the calibration and calculate the expected position for the 1/3 filling, which aligns well with the observed 1/3 state. Note that the 1/3 state shows a plateau instead of a sharp minimum because it has a very high resistivity such that the MIM signal is saturated at the lower limit (see the MIM response curve in Fig. 1b in the main text). If we use the center of the plateau to identify the actual position for the 1/3 state, the deviation between the center and the calculated 1/3 position is only $\delta n = 0.023$, most of which should come from the low gate voltage range near 1/3. The filling fractions for other observed states are all above 1/2, so the deviation from the linear relation should be much smaller.

Table S2 lists the filling values calculated from the observed minima and compare with the assigned fractional values. For most states, the deviation is less than 0.0050. The voltage step used in our measurement is 20 mV, corresponding to a filling value of 0.0063. Most deviations are within one data point. The few states with larger deviations are ±5/6, 5/9 and 3/4. For each of these states, we find the best alternative fractional value that match the observed filling very well. To check how the ordering pattern changes, we have performed simulations for these alternative filling fractions. As illustrated in the Fig. S14, compare the 2/11 vs 1/6 patterns and the 5/21 vs 1/4 patterns, the alternative patterns can be considered as the domain states of the original patterns. The formation of domain boundaries allows to accommodate extra or missing charges, so that the ordered pattern can be stabilized over a finite range of density around the nominal fractional value. Similar behavior has been demonstrated by a previous study on the charge ordering in a square lattice[1]. In the experiment, this behavior will lead to a broadening of the dips in the MIM signal. In our data, the simpler fractions fall well within the dips and only have small deviations from the local minima, therefore, we assign the simpler fractions to these features.

For the 5/9 state, the alternative filling at 6/11 shows a different stripe pattern. Our simple Coulomb gas model does not produce a robust ground state for 5/9, probably due to details that are not captured by our model. But based on the above analysis, we believe a simpler fraction should still better represent the state.

**References:**

1. Rademaker, L., Pramudya, Y., Zaanen, J. & Dobrosavljević, V. Influence of long-range interactions on charge ordering phenomena on a square lattice. *Phys. Rev. E* **88**, 032121 (2013).

| Assigned fraction | Observed filling | Deviation | Best alternative |
|---|---|---|---|
| -8/9 (-0.8889) | -0.8866 | 0.0022 | |
| -5/6 (-0.8333) | -0.8182 | 0.0151 | -9/11 (-0.8182) |
| -7/9 (-0.7778) | -0.7795 | 0.0017 | |
| 1/2 (0.5) | 0.5036 | 0.0036 | |
| 5/9 (0.5556) | 0.5433 | 0.0123 | 6/11 (0.5455) |
| 3/4 (0.75) | 0.7582 | 0.0082 | 16/21 (0.7619) |
| 5/6 (0.8333) | 0.8150 | 0.0183 | 9/11 (0.8182) |
| 6/7 (0.8571) | 0.8530 | 0.0042 | |

Table S2. The experimentally determined filling values compared with the assigned fractions.

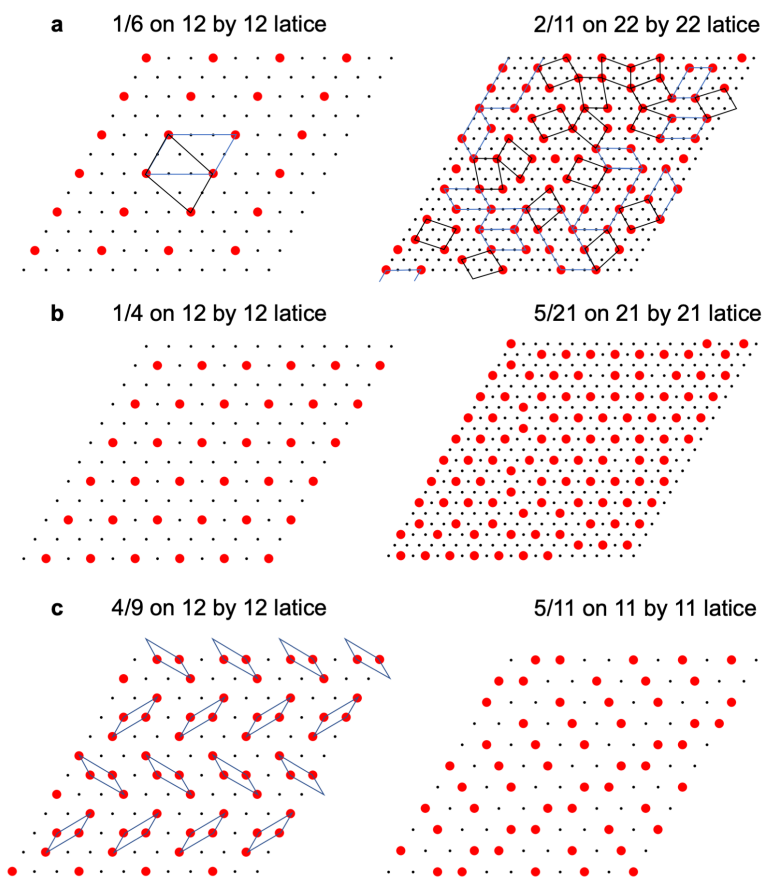

Figure S14. Simulated patterns for (a) 1/6 vs 2/11, the blue and black parallelograms indicate two equivalent unit cells in the 1/6 state; (b) 1/4 vs 5/21; (c) 4/9 vs 5/11.

13